\documentclass[prd,twocolumn,amsmath,showpacs,nofootinbib]{revtex4-1}
\usepackage{graphicx}
\usepackage{dcolumn}
\usepackage{bm}
\usepackage{ulem}

\usepackage[usenames,dvipsnames]{color}
\definecolor{mightnightblue}{RGB}{25,25,112}

\definecolor{brown}{rgb}{0.59, 0.29, 0.0}

\def\vev#1{\left\langle #1\right\rangle}

\def\21{$\mathrm{SU(2)_L \otimes U(1)_Y}$}

\def\sm{standard model }

\begin{document}
\newcommand{\hs}{\hspace*{0.5cm}}
\newcommand{\vs}{\vspace*{0.5cm}}
\newcommand{\be}{\begin{equation}}
\newcommand{\ee}{\end{equation}}
\newcommand{\bea}{\begin{eqnarray}}
\newcommand{\eea}{\end{eqnarray}}
\newcommand{\ben}{\begin{enumerate}}
\newcommand{\een}{\end{enumerate}}
\newcommand{\bwt}{\begin{widetext}}
\newcommand{\ewt}{\end{widetext}}
\newcommand{\nn}{\nonumber}
\newcommand{\crn}{\nonumber \\}
\newcommand{\Tr}{\mathrm{Tr}}
\newcommand{\non}{\nonumber}
\newcommand{\noi}{\noindent}
\newcommand{\al}{\alpha}
\newcommand{\la}{\lambda}
\newcommand{\bet}{\beta}
\newcommand{\ga}{\gamma}
\newcommand{\va}{\varphi}
\newcommand{\om}{\omega}
\newcommand{\pa}{\partial}
\newcommand{\+}{\dagger}
\newcommand{\fr}{\frac}
\newcommand{\bc}{\begin{center}}
\newcommand{\ec}{\end{center}}
\newcommand{\Ga}{\Gamma}
\newcommand{\de}{\delta}
\newcommand{\De}{\Delta}
\newcommand{\ep}{\epsilon}
\newcommand{\varep}{\varepsilon}
\newcommand{\ka}{\kappa}
\newcommand{\La}{\Lambda}
\newcommand{\si}{\sigma}
\newcommand{\Si}{\Sigma}
\newcommand{\ta}{\tau}
\newcommand{\up}{\upsilon}
\newcommand{\Up}{\Upsilon}
\newcommand{\ze}{\zeta}
\newcommand{\ps}{\psi}
\newcommand{\Ps}{\Psi}
\newcommand{\ph}{\phi}
\newcommand{\vph}{\varphi}
\newcommand{\Ph}{\Phi}
\newcommand{\Om}{\Omega}
\title{Asymmetric Dark Matter, Inflation and Leptogenesis from B-L Symmetry Breaking}%

\author{P. V. Dong}\email{pvdong@iop.vast.ac.vn}\affiliation{Institute of Physics, Vietnam Academy of Science and Technology, 10 Dao Tan, Ba Dinh, Hanoi, Vietnam}
\author{D. T. Huong}\email{dthuong@iop.vast.ac.vn}\affiliation{Institute of Physics, Vietnam Academy of Science and Technology, 10 Dao Tan, Ba Dinh, Hanoi, Vietnam}
\author{Daniel A. Camargo}\email{dacamargov@gmail.com}\affiliation{International Institute of Physics, Universidade Federal do Rio Grande do Norte, Campus Universitario, Lagoa Nova, Natal-RN 59078-970, Brazil}
\author{Farinaldo S. Queiroz}\email{farinaldo.queiroz@iip.ufrn.br}\affiliation{International Institute of Physics, Universidade Federal do Rio Grande do Norte, Campus Universitario, Lagoa Nova, Natal-RN 59078-970, Brazil}
\author{Jos\'e W. F. Valle}\email{valle@ific.uv.es} \affiliation{AHEP Group, Instituto de F\'isica Corpuscular – C.S.I.C./Universitat de Valencia
Edificio de Institutos de Paterna, C/Catedratico Jos\'e
Beltran, 2 E-46980 Paterna (Valencia) - SPAIN}

\begin{abstract}

We propose a unified setup for dark matter, inflation and baryon asymmetry generation through the neutrino mass seesaw mechanism. Our scenario emerges naturally from an extended gauge group containing $B-L$ as a non-commutative symmetry, broken by a singlet scalar that also drives inflation. Its decays reheat the universe, producing the lightest right-handed neutrino. Automatic matter parity conservation leads to the stability of an asymmetric dark matter candidate, directly linked to the matter-antimatter asymmetry in the universe. 
\end{abstract}

\date{\today}

\maketitle

\section{Introduction} 

The need to account for neutrino oscillations requires new physics
beyond the standard model. In addition, the cosmological challenges of
particle physics, such as the need to account for dark matter,
inflation and reheating, as well as the matter-antimatter asymmetry of
the Universe, all suggest the existence of new physics.
Traditional proposals, based upon supersymmetry, grand unification or
extra dimensions, address only some of these issues separately. It is
therefore desirable to find a comprehensive theory that can provide
a common framework to address all of these puzzles.\\

The seesaw mechanism \cite{Minkowski:1977sc,GellMann:1980vs,Yanagida:1979as,Glashow:1979nm,Mohapatra:1979ia,Mohapatra:1980yp,Lazarides:1980nt,Schechter:1980gr,Schechter:1981cv} is the most popular way to account for small neutrino masses. Here we assume that neutrino masses arise through the exchange of heavy right-handed neutrinos, hence suppressed by the right-handed mass scale. This also provides an attractive way to understand the observed baryon asymmetry of the Universe through the so-called leptogenesis mechanism \cite{Fukugita:1986hr,Buchmuller:2005eh,Davidson:2008bu}. The latter can be triggered by the CP-violating and out-of-thermal-equilibrium decays of the heavy right-handed neutrinos. These decays are followed by sphaleron processes in the effective standard model. \\

Both the seesaw mechanism and leptogenesis suggest having the $B-L$ charge as a gauge symmetry. Indeed, a gauge completion requires the presence of right-handed neutrinos as basic fermions, due to $B-L$ anomaly cancellation. Moreover, $B-L$ plays a key role in converting the lepton to the baryon asymmetry during the sphaleron phase.\\

In addition, we also require a period of inflationary expansion of the early universe, in order to make the standard Big-Bang picture fully consistent. This phenomenon can be parametrized by the slow-roll time evolution of a scalar field, called inflaton \cite{Starobinsky:1980te,Guth:1980zm,Patrignani:2016xqp}. \\

Last, but not least, the standard model lacks a stable weakly interacting massive relic that can be thermally produced at early times \cite{Jungman:1995df,Bertone:2004pz}.  Moreover, the fact that dark matter searches have yielded null results, leaves the origin of dark matter as a big challenge \cite{Iocco:2015xga,Acharya:2017ttl,DeAngelis:2017gra,Arcadi:2017kky}. Here we propose that, since dark matter and normal matter were thermally connected to begin with, they should be manifestly unified within gauge multiplets. \\

In this paper we show how a $B-L$ gauge extension of the standard model yields a viable dark matter candidate, stabilized in a natural way by the residual matter parity which results from the extended gauge symmetry. This also fits nicely with the neutrino mass generation through the seesaw mechanism. \\

The latter is manifestly unified with inflation, since the inflaton $S$ arises from the field responsible for $B-L$ breaking. This takes place at a scale $\La$. Right-handed neutrinos are produced via inflaton decays during reheating. As a result, in our $B-L$ gauge theory setup we have that neutrino mass generation, inflation and reheating as well as leptogenesis are all mutually interconnected. In addition, our $B-L$ theory naturally possesses matter parity as a residual gauge symmetry. These features are in sharp contrast with the simplest Abelian $B-L$ extensions of the standard model \cite{Montero:2011jk,Schmitz:2012kaa,Sanchez-Vega:2014rka,Rodejohann:2015lca}. \\

In order to demonstrate all these points explicitly we start in Section~\ref{model} by describing a gauge theory that manifestly unifies the $B-L$ and electroweak charges within a 3-3-1-1 scenario in a non-trivial way. In Sect.~\ref{neutrino} we describe neutrino mass generation, while in Sect.~\ref{cosmo} we examine the novel cosmological implications of our scheme concerning the issues of dark matter, inflation and the baryon asymmetry. We conclude in Sect.~\ref{conclusion}.\\
                                                           
\section{\label{model} Non-commutative $B-L$ dynamics}

Without loss of generality, we consider only the simplest scenario, based on the 3-3-1-1 extension of the Glashow-Weinberg-Salam theory \cite{Dong:2013wca,Dong:2014wsa,Huong:2015dwa,Huong:2016ybt,Alves:2016fqe,Dong:2015yra,Dong:2015jxa}. The non-commutative $B-L$ gauge symmetry mechanism we propose is, however, more general. \\

The $SU(3)_L$ symmetry is a direct extension from the $SU(2)_L$ weak isospin. This extended electroweak gauge symmetry is motivated by its ability to predict the number of generations (as being equal to that of colors) as a result of $[SU(3)_L]^3$ anomaly cancellation \citep{Singer:1980sw}.\\

In addition, notice that, like the electric charge $Q$, the $B-L$ charge neither commutes nor closes algebraically within $SU(3)_L$. Hence, in order to get a consistent closed gauge structure, two new Abelian $U(1)_X$ and $U(1)_N$ gauge groups are required, where
\bea 
&& Q=T_3+\beta T_8+X,\\
&& B-L=\beta' T_8 +N.
\eea 
This way we are led to the $SU(3)_C\otimes SU(3)_L\otimes U(1)_X\otimes U(1)_N$ (or 3-3-1-1) group structure. Here the $T_i\ (i=1,2,3,...,8)$ are the $SU(3)_L$ generators, while $X$, and $N$ are associated to $U(1)_X$, and $U(1)_N$, respectively. The parameters $\beta$ and $\beta'$ are embedding coefficients, arbitrary on theoretical grounds, and independent of all anomalies. \
\
The new feature is that, in contrast to the ordinary $B-L$ symmetry, our $B-L$ is a non-commutative gauge symmetry, analogous to $Q$, non-trivially unified with the weak forces. The non-trivial commutations, 
\bea && \left[Q, T_1\pm i T_2\right] = \pm (T_1\pm i T_2),\\
&& \left[Q, T_4\pm i T_5\right] = \mp q (T_4\pm i T_5),\\
&& \left[Q, T_6\pm i T_7\right] = \mp (1+q) (T_6\pm i T_7),\\
&& \left[B-L, T_4\pm i T_5\right] = \mp (1+n) (T_4\pm i T_5),\\
&&\left[B-L, T_6\pm i T_7\right] = \mp (1+n) (T_6\pm i T_7), \eea 
subsequently define the $Q$ and $B-L$ charges for the new particles via the basic relations $$q\equiv -(1+\sqrt{3}\beta)/2 \mathrm{~~~and~~~} n\equiv-(2+\sqrt{3}\beta')/2,$$ respectively. \\

The simplest fermion sector, free of all gauge anomalies, is given as Table~\ref{tab0}.
\begin{table*}  [h]
\begin{tabular}{ccccc}
\\
\hline\hline
Multiplet & $SU(3)_C$ & $SU(3)_L$ & $U(1)_X$ & $U(1)_N$ \\  
\hline
$\psi_{aL}\equiv 
\left(\begin{array}{c}
\nu_{aL}\\
e_{aL}\\
N_{aL}\end{array}\right)$ &  1 & 3 & $\fr{-1+q}{3}$ & $\fr{-2+n}{3}$\\
 $Q_{\al L}\equiv 
\left(\begin{array}{c}
d_{\al L}\\
-u_{\al L}\\
j_{\al L} \end{array}\right)$  & 3 & $3^*$ & $-\fr{q}{3}$ & $-\fr{n}{3}$ \\
$ Q_{3 L}\equiv 
\left(\begin{array}{c}
u_{3L}\\
d_{3L}\\
j_{3L}\end{array}\right) $ & 3 & 3 & $\fr{1+q}{3}$ & $\fr{2+n}{3}$ \\
$ \nu_{aR}$ & 1 & 1 & 0 & $-1$\\
$e_{aR}$ & 1 & 1 & $-1$ & $-1$\\
$ N_{aR}$ & 1 & 1 & $q$ & $n$\\
$u_{a R}$ & 3 & 1 & $\fr 2 3$ & $\fr 1 3$\\  
$d_{aR}$ & 3 & 1 & $-\fr 1 3$ & $\fr 1 3$ \\ 
$j_{\al R}$ & 3 & 1 & $-\fr 1 3 -q$ & $-\fr 2 3 -n$\\ 
$j_{3R}$ & 3 & 1 & $\fr 2 3 +q$ & $\fr 4 3 +n$\\
\hline \hline
$\eta \equiv
\left(\begin{array}{l}
\eta_1\\
\eta_2\\
\eta_3
\end{array}\right)$ & 1 & 3 & $\fr{q-1}{3}$ & $\fr{n+1}{3}$\\
$\rho \equiv
\left(
\begin{array}{l}
\rho_1\\
\rho_2\\
\rho_3
\end{array}\right)$ & 1 & 3 & $\fr{q+2}{3}$ & $\fr{n+1}{3}$\\
$\chi \equiv

\left(
\begin{array}{l}
\chi_1\\
\chi_2\\
\chi_3
\end{array}\right) $ & 1 & 3 & $-\fr{2q+1}{3}$ & $-\fr 2 3 (n+1)$ \\
$\phi$  & 1 & 1 & 0 & 2 \\
\hline \hline
\end{tabular}
\caption{\label{tab0} Field representation content of the model.} 
\end{table*}         
Notice that the scalar content is necessary for realistic symmetry breaking and mass generation.
Here, $a=1,2,3$ and $\al=1,2$ label the particle families. Table \ref{tab2} gives the $Q$, $B-L$ charges of the component fields.\\    

The electrically-neutral scalars can develop vacuum expectation values (vevs) given by     
\bea && \vev{ \eta } = \fr{1}{\sqrt{2}}\left(
\begin{array}{c}
u \\
0\\
0
\end{array}\right),\hs
\vev{ \rho} =
\fr{1}{\sqrt{2}} \left(
\begin{array}{c}
0\\
v \\
0
\end{array}\right),\\
&& \vev{ \chi} =
\fr{1}{\sqrt{2}} \left(
\begin{array}{c}
0\\
0\\
w
\end{array}\right),\hs \vev{ \phi} = \fr{1}{\sqrt{2}} \La.\eea 
Here the vevs $w,\La$ break the 3-3-1-1 symmetry down to the standard model times matter parity, $W_P=(-1)^{3(B-L)+2s}$ (see below), providing masses to the new particles. On the other hand the vevs
$u,v$ break the standard model symmetry down to $SU(3)_C\otimes U(1)_Q$, producing the ordinary particle masses. \\

For consistency, we impose \be \La\gg w \gg u,v,\ee where the first hierarchy states that the $U(1)_N$ breaking scale is much larger than the $SU(3)_L\otimes U(1)_X$ breaking scale, while the second hierarchy is similar to that of the simplest 3-3-1 model~\cite{Singer:1980sw}, and allows for potentially accessible new phenomena.                 \\

\section{\label{neutrino} Neutrino mass generation}

The above non-commutative $B-L$ dynamics provides a natural seesaw mechanism as a result of gauge symmetry breaking. We start with the implementation of the type-I seesaw mechanism [the type-II seesaw alternative in 3-3-1 models has been considered in~\cite{Valle:2016kyz}.].\\

To analyze this we first consider the gauge symmetry breaking. This is governed by the Higgs potential, which can be separated into $V=V(\phi)+V(\eta,\rho,\chi)+V_{\mathrm{mix}}$, where \\[-.3cm] 
\bea V(\phi) &=& \mu^2_\phi \phi^\dagger \phi+ \la (\phi^\dagger \phi)^2, \crn
V(\eta,\rho,\chi) &=&  \mu^2_\rho \rho^\dagger \rho + \mu^2_\chi \chi^\dagger \chi +\mu^2_\eta \eta^\dagger \eta + (\mu \eta \rho \chi +H.c.)\crn
 && + \la_1 (\rho^\dagger \rho)^2 + \la_2 (\chi^\dagger \chi)^2 + \la_3 (\eta^\dagger \eta)^2 \crn
 &&  +\la_4 (\rho^\dagger \rho)(\chi^\dagger \chi)  +\la_5 (\rho^\dagger \rho)(\eta^\dagger \eta) +\la_6 (\chi^\dagger \chi)(\eta^\dagger \eta)\crn
&& +\la_{7} (\rho^\dagger \chi)(\chi^\dagger \rho) +\la_8 (\rho^\dagger \eta)(\eta^\dagger \rho) +\la_{9} (\chi^\dagger \eta)(\eta^\dagger \chi),\crn
V_{\mathrm{mix}} &=& \la_{10}(\phi^\dagger \phi)(\rho^\dagger \rho)+\la_{11}(\phi^\dagger \phi)(\chi^\dagger \chi)\crn
&&+\la_{12} (\phi^\dagger \phi)(\eta^\dagger\eta),\nn\eea where the  $\mu$-type parameters have mass dimension, while $\la$'s are dimensionless.\\[-.3cm] 

The field $\phi$ obtains a large vev, $\La^2=-\mu^2_\phi/\la$, implied by $V(\phi)$ due to $\mu^2_\phi<0,\ \la>0$. Integrating $\phi$ out, one finds that the effective potential coincides with $V(\eta,\rho,\chi)$ at the leading order. This potential provides two weak scales $u^2,v^2$ proportional to $-\mu^2_{\eta,\rho}>0$ and the scale $w^2$ proportional to $-\mu^2_\chi>0$. This is totally analogous to the situation in the 3-3-1 model. The conditions for having the above vevs amount to imposing $|\mu_\phi|\gg |\mu_{\chi}|\gg |\mu_{\eta,\rho}|$. Like the 3-3-1 model, a consistent Higgs boson mass spectrum can be achieved when the soft-term $\mu^2$ is negative at the 3-3-1 scale, i.e. $\mu^2<0,\ |\mu|\sim |\mu_\chi|$.   \\[-.2cm]

The Yukawa Lagrangian responsible for neutrino mass generation through the seesaw is given as
\be \mathcal{L}\supset h^\nu_{ab}\bar{\psi}_{aL}\eta\nu_{bR} +\fr 1 2 f^\nu_{ab}\bar{\nu}^c_{aR}\nu_{bR}\phi +H.c.\ee 
Note that $\phi = \fr{1}{\sqrt{2}}(\La + S+i A)$ with a nonzero value for the scale $\La$. Since $\phi$ has $N=B-L=2\neq 0$, its vev breaks these charges, providing Majorana masses for $\nu_R$ as well as for the $U(1)_N$ gauge boson at the scale $\La$. Since $\La$ must be substantially larger than the weak scale, the $U(1)$ gauge boson is too heavy for detection. After the electroweak symmetry breaking, one generates the Dirac neutrino mass term via $\eta_1=\fr{1}{\sqrt{2}}(u+S_1+i A_1)$. \\

The total neutrino mass generation Lagrangian is 
\be \mathcal{L}\supset -\fr 1 2 (\bar{\nu}_{aL}\ \bar{\nu}^c_{aR})\left(
\begin{array}{cc}
0 & m_{ab} \\
m_{ba} & M_{ab}
\end{array}\right)\left(\begin{array}{c}
\nu^c_{bL} \\
\nu_{bR}\end{array}\right)+H.c.,\ee where $M=-f^\nu \La/\sqrt{2}$ and $m=-h^\nu u/\sqrt{2}$.  Since $\La\gg u$, the seesaw mechanism yields the observed neutrino ($\sim \nu_L$) masses in the usual manner as 
\be m_\nu = -m M^{-1} m^T= h^\nu (f^\nu)^{-1} (h^\nu)^T \fr{u^2}{\sqrt{2}\La}.\ee 
The heavy neutrinos ($\sim \nu_R$) gain masses at the $B-L$ breaking scale $\La$. In order to obtain $m_\nu \sim 0.1$ eV we assume $\La\sim 10^{14}$ GeV, if $h,f\sim$ 1, since $u\sim$ 100 GeV. This breaking scale is consistent with the inflation scale, as discussed below.    \\[-.3cm]  


Since $B-L=\beta' T_8 + N$ annihilates the non-trivial vacua, $[B-L]\langle \eta,\rho,\chi\rangle =0$, for $u,v,w\neq 0$, it follows that the gauge group $SU(3)_L\otimes U(1)_N$ contains a residual conserved $B-L$ charge, under which a field transforms as \be \Phi \to  \Phi'=U(\om)\Phi,\hs U(\om)=e^{i\om (B-L)}.\ee However, $B-L$ is broken by $\langle \phi\rangle $ since $[B-L]\langle \phi\rangle = \sqrt{2}\La\neq 0$. The remnant of $B-L$ preserves the vacuum, $U(\om)\langle \phi\rangle =\langle \phi\rangle$. We obtain $e^{i\om2}=1$ or $\om=m\pi$ for $m$ integer. The residual transformation is $U(m\pi)=e^{im\pi (B-L)}=(-1)^{m(B-L)}$. Multiplying $U(3\pi)$ with spin parity $(-1)^{2s}$ due to Lorentz symmetry, yields a matter parity $W_P=(-1)^{3(B-L)+2s}$. While this is a commonly known symmetry, in our case it originates as a residual gauge symmetry,
\be W_P=(-1)^{3(\beta' T_8 + N)+2s}, \ee 
which transforms non-trivially the particles with ``wrong'' $B-L$ charges as seen in Tab. \ref{tab2} (thus the label ``$W$''). In other words, since $B-L$ is non-commutative, $W$-parity separates the gauge multiplets into two parts, including normal particles (W-even) and ``wrong'' particles (W-odd), respectively. [In supersymmetry, it separates super-multiplets, by contrast.] One can show that $P^+$ and $P^-$ particles always appear in pairs in interactions. Indeed, assume that an interaction has $x$ $P^+$ fields and $y$ $P^-$ fields. The conservation of $W$-parity implies $(P^+)^x(P^-)^y=1$ which happens only if $x=y$, for arbitrary $x,y$ integers. Thus, the lightest $W$-particle is stable and, if electrically and color neutral, can be responsible for dark matter.\\

The colorless $W$-particles possess electric charges $\pm q,\pm (1+q)$. Hence, we may have two dark matter options, according to whether $q=0$ and $q=-1$, or $\beta=-1/\sqrt{3}$ and $\beta=1/\sqrt{3}$, respectively. 

The model with $q=0$ yields three potential dark matter candidates, $N$, $W'$, and $\eta_3$, whereas the model with $q=-1$ yields two possible dark matter candidates, $\rho_3$, and $W''$. The former has a correspondence to the original 3-3-1 model~\cite{Singer:1980sw}, while the latter does not. All of these candidates have masses proportional to the $w$ scale times the relevant coupling constants. 

In this work, we consider the simplest but nontrivial case, where $q=0$ and $n=0$, hence $\beta'=-2/\sqrt{3}$, which has been extensively studied~\cite{Dong:2013wca,Dong:2014wsa,Dong:2015yra,Huong:2015dwa,Dong:2015jxa,Huong:2016ybt,Alves:2016fqe} under the assumption that the relics of $N$, $\eta_{3}$ (or $\rho_3$) were thermally produced. In such case the vectors, such as $W'$ or $W''$, cannot be viable dark matter candidates, since they annihilate, before freeze-out, into $W$ bosons via gauge self-interactions. In the present work we provide an alternative interpretation for the dark matter abundance, called asymmetric dark matter \cite{Nussinov:1985xr,Zurek:2013wia,Petraki:2013wwa}, where all possible dark matter types, including the vector one, could be viable.

\begin{table*}[h]
\begin{tabular}{lccccccccccccccccccccccccc}
	\hline\hline
	Particle & $\nu_a$ & $e_a$ &   $u_a$   &   $d_a$    & gluon & $\ga$ & $W$ & $Z$ & $Z'$ & $Z''$ & $\eta_{1}$ & $\eta_2$ & $\rho_{1}$ & $\rho_{2}$ & $\chi_3$ & $\phi$ & $N_a$ &   $j_\al $    &     $j_3$     &  $W'$  & $W''$  & $\eta_3$ & $\rho_3$ & $\chi_{1}$ & $\chi_{2}$ \\ \hline
	$Q$      &    0    & $-1$  & $\fr 2 3$ & $-\fr 1 3$ &   0   &   0   & $1$ &  0  &  0   &   0   &    $0$     &   $-1$   &    $1$     &    $0$     &    0     &   0    &  $q$  & $-\fr 1 3 -q$ & $\fr 2 3 + q$ &  $-q$  & $-1-q$ &   $q$    &  $1+q$   &    $-q$    &   $-1-q$   \\
	$B-L$    &  $-1$   & $-1$  & $\fr 1 3$ & $\fr 1 3$  &   0   &   0   &  0  &  0  &  0   &   0   &     0      &    0     &     0      &     0      &    0     &   2    &  $n$  & $-\fr 2 3 -n$ & $\fr 4 3 +n$  & $-1-n$ & $-1-n$ &  $1+n$   &  $1+n$   &   $-1-n$   &   $-1-n$   \\
	$W_P$    &    1    &   1   &     1     &     1      &   1   &   1   &  1  &  1  &  1   &   1   &     1      &    1     &     1      &     1      &    1     &   1    & $P^+$ &     $P^-$     &     $P^+$     & $P^-$  & $P^-$  &  $P^+$   &  $P^+$   &   $P^-$    &   $P^-$    \\ \hline\hline
\end{tabular}
\caption{\label{tab2} $Q$, $B-L$, and $W_P$ values for the model particles, where $P^\pm \equiv (-1)^{\pm(3n+1)}$ are non-trivial for $n\neq \fr{2m-1}{3}$. When $n=\fr{2m}{3}$, $W$-particles become odd, $P^{\pm}=-1$. The antiparticles have opposite $Q$ and $B-L$, while $W_P$ is conjugated, $(P^\pm)^\dagger =P^\mp$.} 
\end{table*}                     

\section{Cosmology in 3-3-1-1 model}
\label{cosmo}

In this section we examine how the non-commutative $B-L$ dynamics provides a natural comprehensive scenario to account for inflation, dark matter, and leptogenesis. This provides a new realization of the idea that inflation and dark matter have as common origin, the neutrino mass seesaw mechanism, proposed in~\cite{Boucenna:2014uma}. \\

Indeed, in the present context, a new superheavy Higgs scalar $S$ which breaks $U(1)_N$ is required, and can behave as an inflaton field, driving the early accelerated expansion of the universe (see below). Inflaton decay only produces superheavy dark matter relics at the very large $\La$ scale \cite{Huong:2016ybt}. Fortunately, it also decays into right-handed neutrinos $\nu_R$, whose decays may yield CP-asymmetric final states consisting of normal matter $\nu_R\to   \eta_2 e$, as well as dark matter, $\nu_R\to  \eta_3 N$. The first mode yields the baryon asymmetry, while the second mode may play the main role in explaining the dark matter asymmetry, both arising from the standard leptogenesis mechanism. \\

 The asymmetric dark matter relics may be either $\eta_3$ or $N$. Here, $\eta_3$ combines the scalar candidate, called $H'$, and the Goldstone of $W'$, called $W'_L$~\cite{Dong:2014wsa}. The present-day dark matter and normal matter relics have as common source the right-handed $\nu_R$ mediating the seesaw mechanism. All candidate types, fermion ($N$), scalar ($H'$), and vector $(W'_L)$, can contribute to the asymmetric dark matter (a detailed evaluation is given below). 

\subsection{\label{inflation} Inflation} 
In this subsection, we consider the inflationary scenario. This is linked to the singlet scalar $\phi$, which breaks the $U(1)_{N}$ symmetry. Chaotic inflation arises from a tree-level scalar potential of the type 
\bea
V(\phi) =\mu^2_\phi \phi^\dag \phi +\la (\phi^\dag \phi)^2,
\label{tree1}\eea
where the scalar $\phi$ couples to additional fields such as the $U(1)_N$ gauge boson ($C$), fermion fields ($\nu_{aR}$), and scalar fields ($\eta,\rho,\chi$). Through quantum corrections, these couplings modify the tree level inflationary potential. We denote the inflaton as $\Phi= \sqrt{2}\Re(\phi)$, since the imaginary part $\Im(\phi)$ is an unphysical Golstone boson that is gauged away\footnote{This is in contrast to the situation considered in~\cite{Boucenna:2014uma}.}. After including one-loop corrections one has \cite{Coleman:1973jx}
\bea
V(\Phi) &=& \fr{\la}{4}(\Phi^2-\La^2)^2+\fr{a}{64 \pi^2}\Phi^4 \ln\fr{\Phi^2}{\Delta^2}\crn
&&+\fr{9\la^2}{64\pi^2}(\Phi^2-\La^2/3)^2\ln\fr{\Phi^2-\La^2/3}{\Delta^2}+V_0,
\label{effphi}\eea
where the renormalization scale $\Delta$ is arbitrary, the parameter $\La$ is defined as $\La^2=-\mu^2_\phi/\la$ (as usual), $V_0$ is the free/vacuum energy, and 
\bea
a&=&-\fr 1 2 \sum_{i=1}^3(f_{ii}^\nu)^4+48g_N^4+\fr{1}{4}(\la^2_{10}+\la_{11}^2+\la_{12}^2), \label{eff}
\eea
where $f^\nu$ is taken to be flavor-diagonal. This potential always contains a consistent local minimum responsible for the $U(1)_N$ breaking with a suitable choice of the parameters, e.g. $a/\la>-63.165$ for $\Delta \geq \La$.    \\

Notice that in Ref.~\cite{Huong:2015dwa} cosmic inflation was studied with the inflationary potential given in (\ref{effphi}), but the role of the free energy $V_0$ was ignored. Hence, the predicted results for the spectral index $n_s$ and tensor-to-scalar ratio $r$ and running index $\al$ were not fully consistent with the latest experimental results from WMAP and Planck~\cite{Hinshaw:2012aka,Ade:2013zuv,Ade:2015lrj}. In \cite{Huong:2016ybt}, we have interpreted $V_0$ for multi/single-field inflationary scenarios in another setup when the 3-3-1 breaking scale is comparable to the $U(1)_N$ scale. Here, we reconsider the original inflationary scenario by including the contribution of $V_0$, consistent with the leptogenesis scenario. \\

As mentioned, the spectral index $n_s$, tensor-to-scalar ratio $r$, and running index $\al$ are related to slow-roll parameters, $\epsilon=\fr 1 2 m^2_P(V'/V)^2$, $\eta=m^2_P V''/V$, and $\zeta^2=m^4_PV'V'''/V^2$, as follows
\be
 n_s\simeq 1-6\epsilon+2\eta,\ r\simeq 16 \epsilon,\ \al\simeq 16 \epsilon \eta-24\epsilon^2-2 \zeta^2,
\ee where $m_P=\sqrt{8\pi G_N}\simeq 2.4\times 10^{18}$ GeV is the reduced Planck mass. 
The curvature perturbation is
\bea
\Delta^2_{\mathcal{R}}=\fr{V}{24 \pi^2 m_P^4 \epsilon}=2.215 \times 10^{-9},
\eea
at the pivot scale $k_0=0.05\ \mathrm{Mpc}^{-1}$. The number of e-folds during inflation is
\bea
N=\fr{1}{\sqrt{2}m_P} \int_{\Phi_e}^{\Phi_0}\fr{d \Phi}{\sqrt{\epsilon}},
\eea
where  $\Phi_e$ denotes the value of $\Phi$ at the end of inflation, i.e. $\epsilon (\Phi_e) \simeq 1$, $\Phi_0$ is the field value at the horizon exit according to $k_0$, and the value of $N$ is around 50--60 depending on the size of the inflation scale. The parameter $\la$ can be appropriately fixed from the $\Delta_{\mathcal{R}}^2$ constraint. Hence, we are left with $r,n_s,\al$ which are given as functions of $\Phi$ varying from $\Phi_0$ to $\Phi_e$, for selected values of the parameters, $a/\la$, $\La$, $\Delta$, and $V_0$. \\  

By assumption, inflation may occur as the inflaton field slowly rolls to the present potential minimum $(\sim \La)$ from the left, $\Phi < \La$. However, this inflationary scenario seems to be excluded \cite{Huong:2015dwa} because the predicted values of $\Delta_{\mathcal{R}}^2$ and $r$ are not in agreement with the WMAP9 \cite{Hinshaw:2012aka} and Planck \cite{Ade:2013zuv,Ade:2015lrj} observations. Hence we assume $\Phi>\La$ during inflation ($\Phi$ slowly 
rolls towards the potential minimum from the right and inflation terminates at the $U(1)_N$ breaking scale). The inflationary potential is governed by the quartic and log terms. In addition, the log terms that are proportional to $\la^2$'s, hence highly suppressed if we impose $\la,\la_{10,11,12} \ll g^2_N, (f_{ii}^\nu)^2$ as required in order to keep the flatness of the inflationary potential. Therefore, the inflationary potential can be rewritten in a simple form as
\be
V(\Phi)\simeq \fr{\la}{4}\left(\Phi^4 +\fr{a^\prime}{64\pi^2} \Phi^4\ln \fr{\Phi^2}{\Delta^2}+V_0^\prime\right),
\label{inflationa}\ee
where $V_0^\prime =\fr{4 }{\la}V_0= \kappa m_P^4$, and   
\be a^\prime =\fr{4}{\la}\left(a+9\la^2\right)\simeq \fr{2}{\la}\left[96g^4_N-\sum_{i=1}^3 (f^\nu_{ii})^4\right].\label{ddh}\ee
That said, $n_s, r, \al$ would constrain $a', \Delta, \kappa$, while $\Delta_{\mathcal{R}}^2$ fixes $\la$ as already mentioned. \\

\begin{figure}[h]
	\centering
	\includegraphics[scale=0.5]{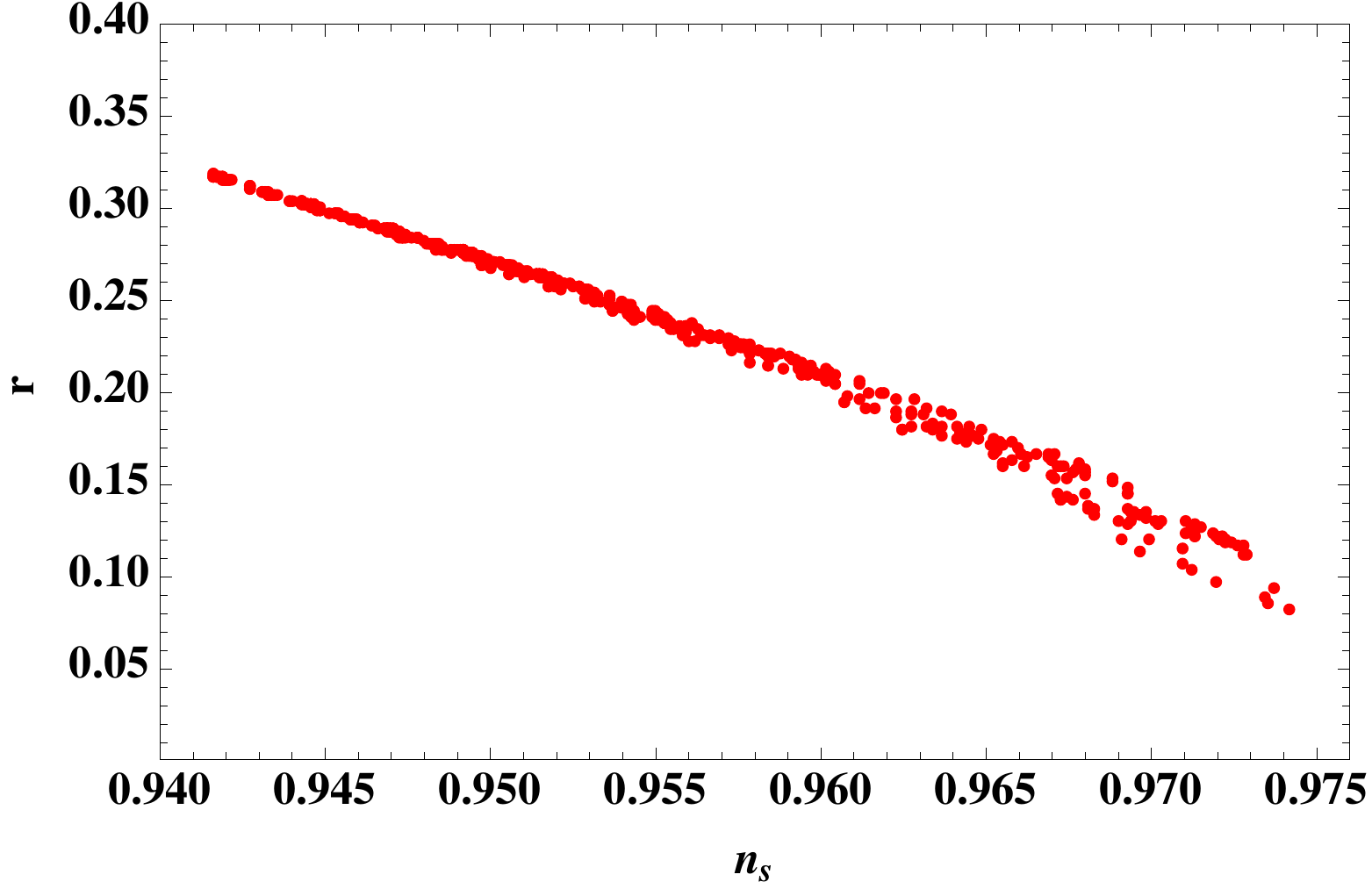}
	\includegraphics[scale=0.5]{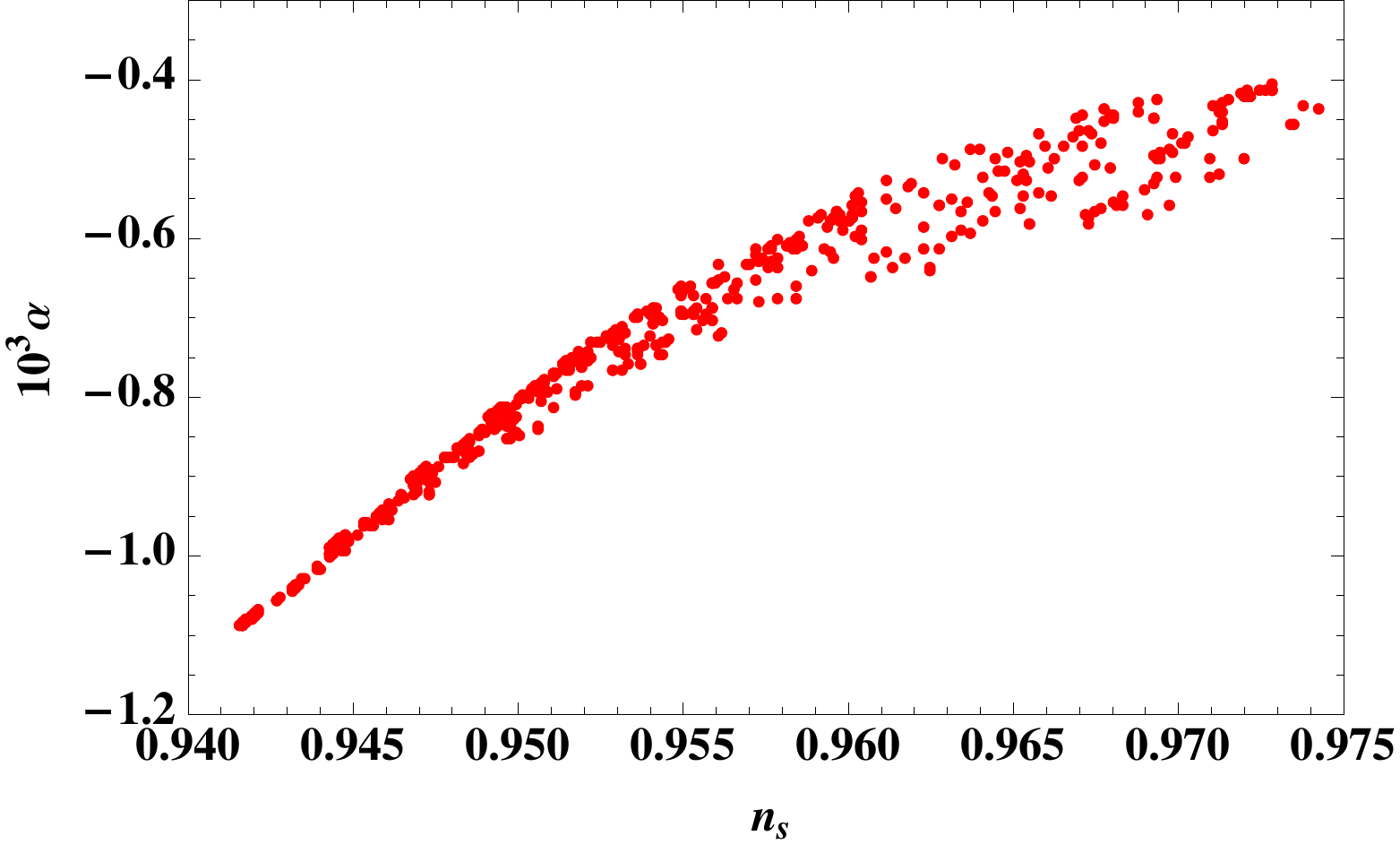}
	\includegraphics[scale=0.5]{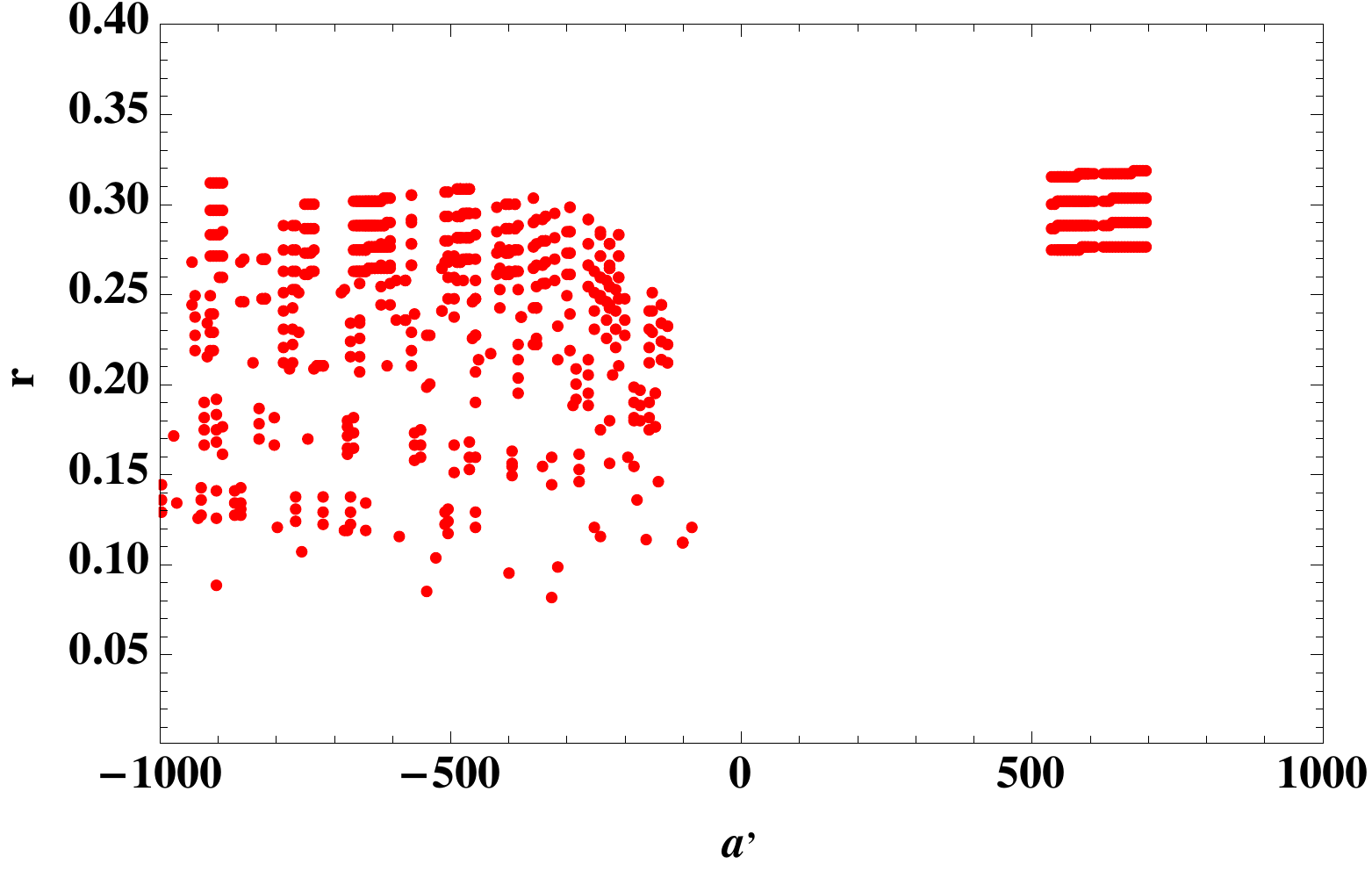}
	\caption{\label{fig1} $r$ vs $n_s$, $\al$ vs $n_s$, and $r$ vs $a^\prime$ achieved according to $a^\prime, \Delta, \kappa $ in the ranges,
		$-10^3<a^\prime<10^3$, $0.1 m_P<\Delta< 120 m_P$, $1<\kappa <10^{8}$, and the number of e-folds in the range $50\le N \le 60$. }
\end{figure}

The spectral index, tensor-to-scalar ratio, and running index are depicted in Fig. \ref{fig1}. Their predicted values are in good agreement with the WMAP9 and Planck results, $n_s = 0.968 \pm 0.006,\ r<0.11,\ \al = - 0.003 \pm 0.007$ \cite{Hinshaw:2012aka,Ade:2013zuv,Ade:2015lrj}, even with the recent data, $r<0.07$ \cite{Ade:2015tva,Array:2015xqh}. To be concrete, Table \ref{Table1} shows consistent inflationary observables $(n_s,r,\al)$ along with viable parameter regions for $(\kappa, a^\prime, \Delta, \la, \Phi)$. It implies that the vacuum energy $V'^{1/4}_0$ varies from a few to hundred $m_P$. The inflaton field is also larger than few tens up to hundreds of the Planck scale, corresponding to the renormalization scale around $\Delta=\mbox{10--100}\ m_P$. Further, we do not find any viable parameter space for $(\kappa, \la, \Delta, a^\prime)$ consistent with the experimental data for $\Delta \lesssim m_P$. 
\begin{table*}
\begin{tabular}{ccccccccc}
	\hline \hline
$\kappa$ & $a^\prime$ & $\Delta[m_P]$ & $\Phi_e[\Delta]$ & $\Phi_0[\Delta]$&$n_s$  &$r$  &$ \al [10^{-3}]$   &$\la[10^{-12}]$ \\ 	\hline 
$10^9$ &$-108$	&$30.1$	&	$7.5$ & $	7.9$ & $	0.975$ &	$0.076$	&$-0.417$ & $0.111$ \\ 
$10^8$ & $-174$	& $120$	&	$2.5$ & 	$2.6$ &	$0.974$	& $0.079$&	$-0.432$&   $0.094$ \\ 
$10^7$ &$-152$ &  $	10$ &	$	6$ &	$7.3$	& $0.974$	& $0.101$ & 	$-0.413$ &	$3.3$ \\ 
$10^6$ & $-99$	&$15.1$& $2.7$ &	$3.3$ & 	$0.966$	& $0.058$ &	$-0.516$& $48.95$ \\ 
$10^5$ & $-212$ & 	$115.1$	& $2.1$	& $2.2$	& $0.971$	& $0.090$ &	$-0.534$ &	0.142  \\ 
$10^4$& $-71$ & $15.1$& $	9.2$&	$10$&	$0.972$ &	$0.098$ &	$-0.495$ &$2.5 $ \\ 
$10^3$ & $-212$	& $115.1$ & $2.1$& $2.2$ &	$0.971$ & 	$0.090$&	$-0.533$ &	$0.150$  \\ 
$10^2$& $-212$&	$115.1$	&	$2.1$&	$2.2$	& $0.971$ &	$0.090$ &	$-0.533$	& $0.151$ \\ \hline\hline
\end{tabular} 
\caption{\label{Table1} The actual parameter regimes that recover all the experimental data \cite{Ade:2013zuv,Ade:2015lrj,Hinshaw:2012aka,Ade:2015tva,Array:2015xqh}.}
\end{table*}

It is clear that the interactions of $\Phi$, which induce a negative contribution to the effective potential ($a'<0$), appropriately fit all inflationary data, and the typical value is taken as $a'\sim -100$. This value satisfies the minimum condition for the full potential in (\ref{effphi}) given that $\La < \Delta$. Thus, it is suitable to take $\La \sim m_P$. Moreover, the scalar self-coupling is typically $\la \sim 10^{-13}$. Thus, we obtain from (\ref{ddh}) as follows
\be \sum_i\left(f^\nu_{ii}/\sqrt{2}\right)^4 -  \fr 3 2 \left(2g_N\right)^4 \sim 10^{-12}.\ee 

Given that  $m_{\nu_R}=-f^\nu\vev{ \Phi}/\sqrt{2}$ and $m_{C}=2g_N\vev{\Phi}$, 
and choosing reasonable values for $g_N$ and $f^\nu_{ii}\sim 10^{-3}$, 
one can assume the heaviest of $\nu_{iR}$ to be slightly heavier than the $U(1)_N$ gauge boson.\\

The vev of $\Phi$ arises from minimizing the scalar potential, $V'=0$, leading to $\vev{\Phi} \sim \La$. The inflaton mass is approximated as 
\be m_\Phi=\sqrt{V''}\sim \sqrt{\la}\La\sim 10^{12}\ \mathrm{GeV},\ee which is much smaller than the $U(1)_N$ gauge boson and largest right-handed neutrino masses, since $\sqrt{\la}\ll g_N,f^\nu_{ii}$ for some $i$. To make the leptogenesis mechanism viable, one assumes hierarchical Yukawa couplings $f^\nu_{11}\sim \sqrt{\la} \ll f^\nu_{22}\sim f^\nu_{33}\sim g_N$. It follows that
\be m_{\nu_{2,3R}}\sim m_C\gg m_\Phi\sim m_{\nu_{1R}}.\label{mkkk}\ee 
The inflaton cannot decay into $U(1)_N$ gauge bosons and right-handed neutrinos $\nu_{2,3R}$. After inflation, the inflaton decays to scalars $(\rho,\eta,\chi)$ or to the lightest right-handed neutrinos $\nu_{1R}$, which reheat the Universe by thermalizing with the 3-3-1 model particles. \\

The decay rates associated to $\Phi \to  \nu_{1R}\nu_{1R}$ and $\Phi \to  \rho^\dagger\rho,\eta^\dagger\eta,\chi^\dagger\chi$ are given, respectively, by 
\be \Ga_{\nu_{1R}}=\fr{(f^\nu_{11})^2m_\Phi}{32\pi}\hs \Ga_{\rho,\eta,\chi}=\fr{\la^2_{10,12,11}\langle \Phi\rangle^2}{16\pi m_\Phi}.\ee 
Successful leptogenesis (see the next section) is viable for both cases, if $\Phi$ dominantly decays into a pair of scalars, while $\nu_{1R}$ is thermally produced, or vice versa. In the first case, we must impose $|\la_{10,11,12}|\gg |f^\nu_{11}|\sqrt{\la}\sim 10^{-13}$. Since $\la_{10,11,12}\ll g^2_N,(f^\nu_{22,33})^2\sim 10^{-6}$, we assume $\la_{10,11,12}\sim 10^{-9}$. The reheating temperature is obtained by $\Ga=\Ga_\rho + \Ga_\eta + \Ga_\chi$ as follows
\bea T_R&=&\left(\fr{90}{\pi^2 g_*}\right)^{1/4}\sqrt{m_P\Ga}\crn
&\simeq&10^{11}\left(\fr{\la_{\mathrm{mix}}}{10^{-9}}\right)\left(\fr{\La}{m_P}\right)^{1/2}\left(\fr{10^{-13}}{\la}\right)^{1/4} \mathrm{GeV},\eea           
where $\la_{\mathrm{mix}}\equiv \sqrt{\la^2_{10}+\la^2_{11}+\la^2_{12}}$, and $g_*=106.75$ is the effective number of degrees of freedom. The predicted reheating temperature is $T_R\sim 10^{11}$ GeV, for typical values of the parameters. Thus our model provides an alternative to Grand Unification, with the proton automatically stable as a result of the gauge symmetry and $W$ parity. Since supersymmetry is not invoked, we avoid the stringent bounds on the reheating temperature~\cite{Ellis:1983ew}. \\

 The lightest right-handed neutrino $\nu_{1R}$ can be thermally generated during reheating, even though its mass is larger than the reheating temperature. Indeed, radiation only dominates the universe when the temperature lies below the reheating temperature. However, the inflaton decay products $\rho,\eta,\chi$ can rapidly thermalize, forming a plasma with a background temperature much higher than the reheating temperature, e.g. $10^3 T_R$. As a result $\nu_{1R}$ can be created by scattering of light states, or by thermalizing of the heavier $C,\nu_{2,3R}$ \cite{Kuzmin:1997jua,Berezinsky:1997hy,Chung:1998rq}. \\

For the second case, we impose $|\la_{10,11,12}|\ll |f^\nu_{11}|\sqrt{\la}\sim 10^{-13}$. The inflaton mainly decays to two $\nu_{1R}$. The reheating temperature can be computed, yielding $T_R\sim 10^8$ GeV, given that $f^\nu_{11}\sim \sqrt{\la}\sim 10^{-6.5}$. One has a non-thermal leptogenesis mechanism as $\nu_{1R}$ are directly produced from inflaton decays.

\subsection{\label{lg} Leptogenesis: normal vs dark matters}

One of the most attractive features of the current model lies in the lepton sector. The right-handed neutrinos are singlets under $SU(3)_L \otimes U(1)_X$, but transform non-trivially under $U(1)_N$. Since they carry one unit of $B-L$, they acquire Majorana mass (two units of lepton number) due to $U(1)_N$ breaking. This constitutes a source for lepton and dark matter asymmetries in the model. \\[-.2cm] 

The relevant Lagrangian is given by 
\bea \mathcal{L} &\supset& -\bar{e}_{aL}(m_e)_{ab}e_{bR}-\bar{N}_{aL}(m_N)_{ab}N_{bR} -\fr 1 2 \bar{\nu}^c_{aR} M_{ab}\nu_{bR}\crn
&&+h^\nu_{ab}(\bar{e}_{aL}\eta^-_2+\bar{N}_{aL}\eta^0_3)\nu_{bR}+H.c.,\label{qtqt}\eea where the mass matrices, $m_e=-h^e v/\sqrt{2}$ and $m_N=-h^N w/\sqrt{2}$, arise from the Yukawa interactions, $h^e_{ab}\bar{\psi}_{aL}\rho e_{bR}+h^N_{ab}\bar{\psi}_{aL}\chi N_{bR}$, respectively \cite{Dong:2015yra}, while the other terms come from the above seesaw mechanism.  The gauge states $(_a)$ are related to the mass eigenstates~$(_i)$ by mixing matrices, $e_{aL,R}=(V_{eL,R})_{ai}e_{i L,R}$, $N_{aL,R}=(V_{N L,R})_{ai}N_{iL,R}$, and $\nu_{aR}=(V_{\nu R})_{a i} \nu_{i R}$, which so that $V^\dagger_{eL}m_e V_{eR}=\mathrm{diag}(m_e,m_\mu,m_\tau)$, $V^\dagger_{NL} m_N V_{NR}=\mathrm{diag}(m_{N_1},m_{N_2},m_{N_3})$, and $V^T_{\nu R}M V_{\nu R}=\mathrm{diag}(M_1,M_2,M_3)$, respectively, leading to
\bea \mathcal{L} &\supset& \cdots-\fr 1 2 M_i \bar{\nu}^c_{iR}\nu_{iR}+x_{ij} \bar{e}_{iL}\eta^-_2\nu_{jR}\crn
&&+y_{ij}\bar{N}_{iL}\eta^0_3\nu_{jR} + H.c.\label{feyrule}\eea The couplings $x=V^\dagger_{eL}h^\nu V_{\nu R}$ and $y=V^\dagger_{NL}h^\nu V_{\nu R}$ are generally distinct and complex, and hence sources of CP-violation. In addition, we have $x=V'y$ and $x^\dagger x =y^\dagger y$, where $V'=V^\dagger_{eL}V_{NL}$ plays a role similar to the ordinary lepton and quark mixing matrices. \\ [-.2cm]          

\begin{figure}[h]
\includegraphics[scale=0.8]{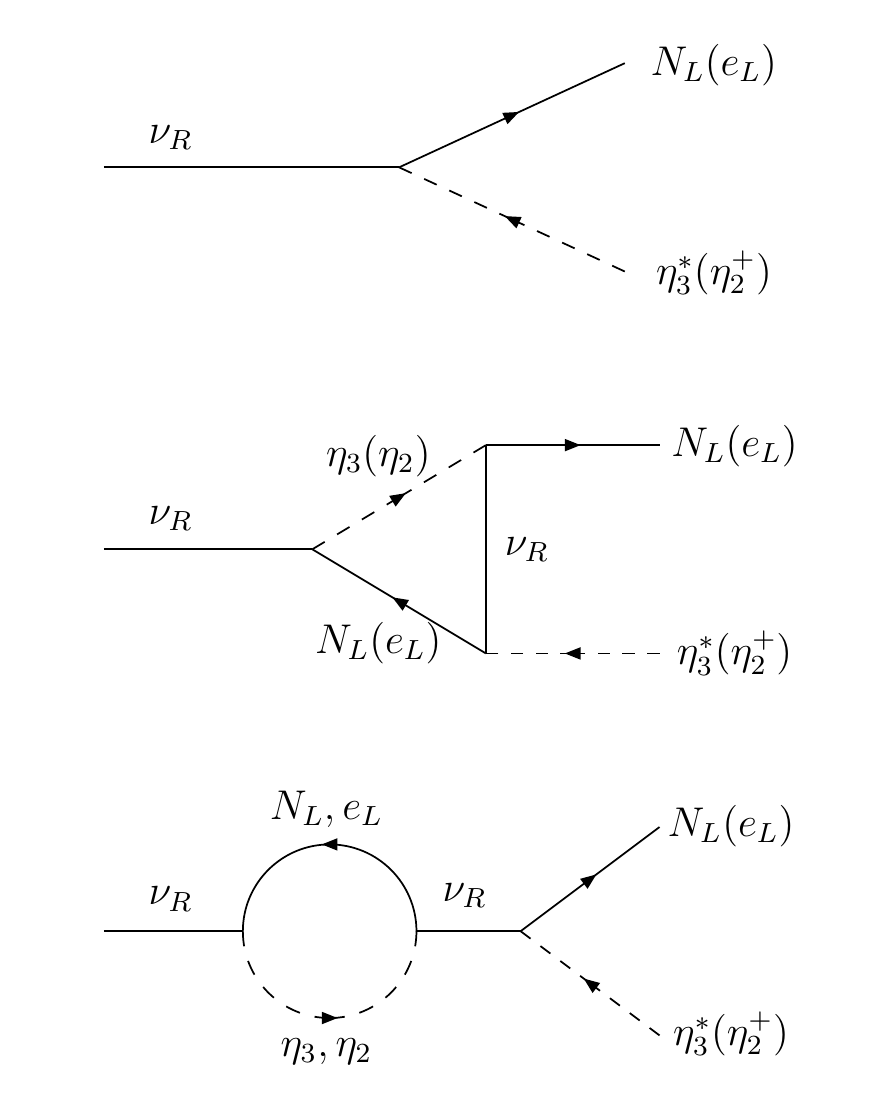}
\caption{\label{leptodig} CP-asymmetric decays of $\nu_R$ into dark matter ($N\eta_3$) and normal matter ($e\eta_2$), respectively, where the Feynman rules and flavor indices can be extracted from (\ref{feyrule}).} 
\end{figure}

Notice that the right-handed neutrinos can decay (out of thermal equilibrium) simultaneously into
\begin{itemize}
\item normal matter: $\nu_{kR}\to   \eta_2 e_i$,
\item dark matter: $\nu_{kR}\to  \eta_3 N_i$,	
\end{itemize}
which subsequently, due to the $W$-parity conservation, generate two different and unrelated CP-asymmetries, $\epsilon_{SM}^{i}$ and $\epsilon_{DM}^{i}$, given respectively  as
\bea
\epsilon_{SM}^{i} &=&\fr{\Gamma(\nu_{kR}\to   \eta_2 e_i)-\Gamma(\nu_{kR}\to   \overline{\eta}_2 \overline{e}_i)}{\Gamma_{\nu_{kR}}}, \\
\epsilon_{DM}^{i}&=& \fr{\Gamma (\nu_{kR}\to  \eta_3 N_i)- \Gamma (\nu_{kR}\to  \overline{\eta}_3 \overline{N_i})}{\Gamma_{\nu_{kR}}}, 
\eea via the Feynman diagrams as depicted in Fig. \ref{leptodig}, where $\Ga_{\nu_{kR}}$ is the total width of $\nu_{kR}$, with assuming that $M_k<M_j$ for all $j\neq k$ (for a fixed $k$, and often chosen $k=1$ as in the previous section). 
The dark matter production $(\eta_3 N)$ is a new observation of this work. Furthermore, it is checked that all other new particles including $W''$ negligibly contribute to the CP-asymmetries, which contrast with \cite{Huong:2015dwa}. We thus obtain 
\bea \epsilon_{SM}^i &=& \fr{1}{16\pi (x^\dagger x)_{kk}}\sum_{j\neq k} \Im [(x^\dagger x)_{jk} x^*_{ij} x_{ik}]g(\xi_{jk}), \\
\epsilon^i_{DM} &=& \fr{1}{16\pi (y^\dagger y)_{kk}}\sum_{j\neq k} \Im [(y^\dagger y)_{jk} y^*_{ij} y_{ik}]g(\xi_{jk}), \eea where $\xi_{jk}=M^2_j/M^2_k$, and \be g(\xi)= \sqrt{\xi}\left[\fr{2}{1-\xi}+1-(1+\xi)\ln\fr{1+\xi}{\xi}\right].\ee 

We stress that the ordinary leptons ($e_i$) each carry a distinct flavor number, $L_i(e_i)=1$, and the CP asymmetries $\epsilon_{SM}^{i}$ are often thought to depend on flavor \cite{Nardi:2006fx}, i.e. each of them creating a separate contribution to the baryon asymmetry, $\eta_{SM}^i$, via the electroweak sphaleron. However, this flavor effect does not hold here since the largest interaction rate corresponding to the tau flavor is $\Ga_\tau \simeq 5\times 10^{-3}(h^\tau)^2 T$, is still slower than the cosmological expansion rate ($H$) for $T= M_k\sim 10^{12}$ GeV, which is just $m_{\nu_{1R}}$ in (\ref{mkkk})~\cite{Cline:1993bd}. The lepton asymmetry has no knowledge of flavor and the net contribution is simply summed as follows 
\bea \epsilon_{SM} &=& \sum_i \epsilon^i_{SM}\crn
&=&\fr{1}{16\pi (h^{\nu\dagger}h^\nu)_{kk}}\sum_{j\neq k} \Im[(h^{\nu\dagger}h^\nu)_{jk}^2] g(\xi_{jk}),\label{vddd}\eea where it is sufficient for us, for simplicity, to take $V_{\nu R}=1$. \\

However, the ``wrong'' particles $N_i$ have $h^{N_i}$ as Yukawa couplings to the new Higgs $\chi$, and we assume $h^{N_i}\gg h^\tau$ for some $i$. Thus, the interaction rate of $N_i$, $\Ga_{N_i} \simeq 5\times 10^{-3}(h^{N_i})^2 T$, is much faster than the Hubble rate during the time of the $N_i$ asymmetry production, $T=M_k\sim 10^{12}$ GeV. In other words, flavor effects should be taken into account for the $N_i$ asymmetric relic. Considering the number of active flavors $n_f=2$, i.e. $h^{N_{1,2}}\sim h^\tau\ll h^{N_3}$, the dark matter asymmetry is given by $\epsilon_{DM}=\epsilon^3_{DM}+\epsilon^{2'}_{DM}$, where the flavor washout factors approach unit for the large $h^\nu$ couplings, and $2'$ is some flavor combination of $i=1,2$ orthogonal to 3. Taking $2'=2$, we get, without loss of generality,   
\bwt \bea \epsilon_{DM} &=& \fr{1}{16\pi (h^{\nu\dagger}h^\nu)_{kk}}\sum_{i=2,3}\sum_{j\neq k}\sum_{l,l'=1,2,3} \Im[(h^{\nu\dagger}h^\nu)_{jk} h^{\nu *}_{lj} h^\nu_{l'k}V^*_{l'i} V_{li} ] g(\xi_{jk})\crn
&=&2\epsilon_{SM}+\fr{1}{16\pi (h^{\nu\dagger}h^\nu)_{kk}}\sum_{i=2,3}\sum_{j\neq k}\sum_{l\neq l'} \Im[(h^{\nu\dagger}h^\nu)_{jk} h^{\nu *}_{lj} h^\nu_{l'k}V^*_{l'i} V_{li} ] g(\xi_{jk}),\label{vhh} \eea\ewt for fixed $k$ and $V\equiv V_{NL}$. It is clear that the dark matter asymmetry recovers the unflavored standard model values for $l=l'$, and the $l\neq l'$ terms are also of the same order. We conclude that $\epsilon_{DM}\sim \epsilon_{SM}$, i.e.  the flavor effect only separates the normal and dark matter asymmetries.  Remark:  When the flavor effects are neglected, i.e. $h^{N_i}\lesssim h^\tau$ for all $i$, the dark matter asymmetry is thus summed over several favors, $\epsilon_{DM}=\sum_{i=1,2,3}\epsilon^i_{DM}=\epsilon_{SM}$, which is the same as the lepton asymmetry. This case also applies for the $\eta_3$ production when it is the lightest $W$-particle.               \\

However, the ``wrong'' particles $N_i$ and $\eta_3$ are singlets under the standard model symmetry. Thus, the CP asymmetries $\epsilon_{DM}^{i}$ are not affected by the electroweak sphaleron nor do they contribute to the baryon asymmetry, as ensured by $W$-parity conservation. The Boltzmann equations can be split into two, one given by the conventional computation for the lepton asymmetry $\epsilon_{SM}=\sum_{i=1,2,3} \epsilon^i_{SM}$ responsible for the baryon asymmetry while the other, given as $\epsilon_{DM}=\sum_{i=2,3}\epsilon_{DM}^{i}$, is responsible for the dark matter asymmetry ($\eta_{DM}$). (Here one assumes $N_i$ to be lighter than $\eta_3$, the inverse case is briefly discussed below). \\

Therefore, the total matter asymmetry of the universe originating from leptogenesis, contains several asymmetries as 
\be \eta_M=\eta_B+\eta_{DM},\ee
where 
\be \eta_{B}=-\fr{8}{15} \sum \eta_{SM}. \ee 
As analyzed, the sphaleron converts only ordinary leptons to ordinary baryons, and this does not work for dark matter, since $L(N)=0$.  The latter also holds for $L(N)\neq 0$ (i.e., $n\neq 0$) i.e. $\eta_3$ is the lightest $W$-particle, because in this case heavier $W$-particles such as exotic quarks ($j_i$) or new gauge and scalar fields ($W''$, $\chi_2$, $\rho_3$) may be created-converted from the $N$ or $\eta_3$ asymmetries via $SU(3)_L$ sphaleron processes that preserve $W$-parity. However, they will decay back to the dark matter, since there is no way to keep them stable, in contrast to the case of ordinary baryons. The total contribution of the two decay modes allows us to explain successfully the baryon and dark matter relics through thermal leptogenesis, as shown below.  \\

The numerical investigation for $\eta_{SM}$ vs $\eta_{DM}$ given in Eqs.~(\ref{vddd}) and (\ref{vhh}) for various choices of $V_{li}$ in the region $5 \times 10^{-11}< \eta_B <10^{-10}$ always yields that the asymmetries in the two decay channels end up of the same order, $\eta_{SM}\sim \eta_{DM}$. On the other hand, the asymmetries are required to reproduce the observed baryon and dark matter abundance. The ratio of the dark matter and baryon density $\Omega_{DM}/\Omega_{B}$ is proportional to that of the asymmetries, $\Omega_{DM}/\Omega_{B}=\eta_{DM}m_{DM}/(\eta_{B}m_p)$ \cite{Graesser:2011wi,Falkowski:2011xh}. Since $\eta_{DM}\sim \eta_B$, the dark matter mass is $m_{DM}\sim m_p$ (the proton mass), so as to fit the observed ratio $\Omega_{DM}\sim5\Omega_{B}$. \\
   
There may be a case when $\nu_{kR}$ is produced directly from the inflaton decay $\Phi \to  \nu_{kR}\nu_{kR}$, with the matter asymmetries followed by a non-thermal leptogenesis. The total CP asymmetry is simply the sum over all flavor CP asymmetries, $\epsilon_{SM}= \sum_{i=1,2,3} \epsilon_{SM}^{i}$ and $\epsilon_{DM}=\sum_{i=1,2,3}\epsilon_{DM}^{i}$, which yields $\epsilon_{SM}=\epsilon_{DM}$. The lepton and dark matter asymmetries are related to the CP asymmetries by $\eta_{L,DM}=\fr{3}{2}\epsilon_{SM,DM}\times Br(\Phi \to  \nu_{kR}\nu_{kR})\times \fr{T_R}{m_{\Phi}}$, respectively, which leads to $\eta_L=\eta_{DM}$. In order to fit $5\simeq \Omega_{DM}/\Omega_B=m_{DM} \eta_{DM}/(m_p \eta_B)\sim m_{DM}/m_p$ one finds that the dark matter relic in the non-thermal case is also light, as above. \\

Two remarks are in order: 

\ben \item If $N$ is an asymmetric dark matter, its mass is close to $m_p$, yet it can avoid restrictions from electroweak precision tests as well as direct searches because, in contrast to Refs.~\cite{Dong:2013wca,Dong:2014wsa,Alves:2016fqe}, it is a standard model singlet, and has only interaction with $Z',Z''$.   
\item If $\eta_3$ is an asymmetric dark matter, we write it in the physical basis as $\eta_3=(wH'-uW'_L)/\sqrt{w^2+u^2}$, where $W'_L\equiv G^*_{X}$~\cite{Dong:2014wsa}. The interaction of $\eta_3$ with the leptons (\ref{qtqt}) yields \bea && \fr{w}{\sqrt{u^2+w^2}}h^\nu_{ab}\bar{N}_{aL}\nu_{bR} H' 
\crn
&&-\fr{u}{\sqrt{u^2+w^2}}h^\nu_{ab}\bar{N}_{aL}\nu_{bR} W'_L.\nn\eea  Since $u\ll w$, the first term generates a light asymmetric dark matter candidate $H'$ which again, thanks to its standard model singlet nature, can avoid all experimental bounds.  However, the second is also consistent with $W'_L$, the Goldstone of $W'$, as dark matter candidate. In this case it implies $\epsilon_{DM}\sim (u/w)^2\epsilon_{SM}$, leading to $m_{W'_L}\sim (w/u)^2m_p\sim 2.5$ TeV, provided that $u\sim 100$ GeV and $w\sim 5$ TeV. This result agrees with~\cite{Falkowski:2011xh}.      
\een

We conclude that in our scenario the asymmetric dark matter may be a light fermion or scalar state ($m_{N,H'}\sim$ GeV) transforming as standard model singlet, or a heavy vector state ($m_{W'_L}\sim$ 2.5 TeV) transforming as a standard model doublet. The dark matter is produced by the standard thermal or non-thermal leptogenesis mechanism. We emphasize that our dark matter phenomenology would then significantly differ from the previous ones discussed in the literature \cite{Cogollo:2014jia,Kelso:2014qka,Alves:2015mua}.           
                
\section{\label{conclusion} Conclusion} 

We have pointed out that the seesaw scenario with non-commuting $B-L$ dynamics can successfully address several of the leading cosmological challenges of the standard model. Our proposal provides a common theoretical setup for the generation of neutrino mass, dark matter, inflation, and baryon asymmetry, from first principles. The seesaw mechanism is based on the non-commutative $B-L$ gauge symmetry present in a 3-3-1-1 \sm extension.  The latter implies a conserved matter parity that stabilizes dark matter, which is manifestly unified with normal matter. Inflation is driven by the $B-L$ breaking field, with appropriate $\La$ scale and $g_N$ coupling. On the other hand, leptogenesis consistently generates not only the present-day baryon asymmetry, but also the dark-matter. 

The model can harbour three asymmetric dark matter candidates: a scalar ($H'$) and a fermion ($N$), both with mass similar to the proton mass, and a vector ($W'$) with mass at the TeV scale. Although strange {\sl a priori}, we note that the
restrictions coming from electroweak precision studies as well as those arising from collider and direct dark matter searches can be avoided in the former case, while TeV masses do indeed match the model parameters expected in the second case.

Note also that our scenario may be potentially viable in other gauge groups containing non-commutative $B-L$, such as $SU(3)_C\otimes SU(P)_L\otimes U(1)_X\otimes U(1)_N$ for $P\geq 3$, and $SU(3)_C\otimes SU(N)_L\otimes SU(M)_R\otimes U(1)_X$ for $(N,M)=(2,3),\ (3,2),\ (3,3),\cdots$. 
Here, the first model class yields a consistent inflation scenario due to the small $g_N$, whereas the latter ones provide successful inflation schemes only if the inflaton, which breaks $B-L$ with a large strength, couples non-minimally to gravity. Likewise, the procedure should hold also for $SO(10)$ and trinification models. 
 
\section*{Acknowledgments}

This work is funded by the Vietnam National Foundation for Science and Technology Development (NAFOSTED) under grant number 103.01-2016.77. DC and FSQ are funded by UFRN and MEC.  FSQ also acknowledges financial support from ICTP-SAIFR FAPESP grant 2016/01343-7.
JWFV is funded by Spanish grants FPA2017-85216-P,  SEV-2014-0398 (MINECO) and  PROMETEOII/2014/084  (Generalitat  Valenciana).

\bibliography{combined2}

 \end{document}